\documentstyle[11pt,amssymb,epsf,cite]{article}

\makeatletter
\@addtoreset{equation}{section}
\makeatother

\def\ben{\begin{equation}}
\def\een{\end{equation}}

  \let\n=\nu

\let\C=\Chi

\def\nn{\nonumber} \def\bd{\begin{document}} \def\ed{\end{document}}
\def\ds{\documentstyle} \let\fr=\frac \let\bl=\bigl \let\br=\bigr
\let\Br=\Bigr \let\Bl=\Bigl
\let\bm=\bibitem
\let\na=\nabla
\let\pa=\partial \let\ov=\overline
\newcommand{\be}{\begin{equation}}
\newcommand{\ee}{\end{equation}}
\def\ba{\begin{array}}
\def\ea{\end{array}}
\def\ft#1#2{{\textstyle{{\scriptstyle #1}\over {\scriptstyle #2}}}}
\def\fft#1#2{{#1 \over #2}}
\def\del{\partial}
\def\vp{\varphi}
\def\sst#1{{\scriptscriptstyle #1}}
\def\oneone{\rlap 1\mkern4mu{\rm l}}
\def\td{\tilde}
\def\wtd{\widetilde}
\def\ie{\rm i.e.\ }
\def\dalemb#1#2{{\vbox{\hrule height .#2pt
        \hbox{\vrule width.#2pt height#1pt \kern#1pt
                \vrule width.#2pt}
        \hrule height.#2pt}}}
\def\square{\mathord{\dalemb{6.8}{7}\hbox{\hskip1pt}}}
\newcommand{\ho}[1]{$\, ^{#1}$}
\newcommand{\hoch}[1]{$\, ^{#1}$}
\newcommand{\bea}{\begin{eqnarray}}
\newcommand{\eea}{\end{eqnarray}}
\newcommand{\ra}{\rightarrow}
\newcommand{\lra}{\longrightarrow}
\newcommand{\Lra}{\Leftrightarrow}
\newcommand{\ap}{\alpha^\prime}
\newcommand{\bp}{\tilde \beta^\prime}
\newcommand{\tr}{{\rm tr} }
\newcommand{\Tr}{{\rm Tr} }
\def\0{{\sst{(0)}}}
\def\1{{\sst{(1)}}}
\def\2{{\sst{(2)}}}
\def\3{{\sst{(3)}}}
\def\4{{\sst{(4)}}}
\def\5{{\sst{(5)}}}
\def\6{{\sst{(6)}}}
\def\7{{\sst{(7)}}}
\def\8{{\sst{(8)}}}
\def\n{{\sst{(n)}}}
\def\cA{{{\cal A}}}
\def\cF{{{\cal F}}}
\def\tV{\widetilde V}
\def\tW{\widetilde W}
\def\tH{\widetilde H}
\def\tE{\widetilde E}
\def\tF{\widetilde F}
\def\tA{\widetilde A}
\def\im{{{\rm i}}}
\def\tY{{{\wtd Y}}}
\def\ep{{\epsilon}}
\def\vep{{\varepsilon}}
\def\R{\rlap{\rm I}\mkern3mu{\rm R}}
\def\bD{{{\bar D}}}

\def\R{\rlap{\rm I}\mkern3mu{\rm R}}
\def\bD{{{\bar D}}}
\def\R{{{\Bbb R}}}
\def\C{{{\Bbb C}}}
\def\H{{{\Bbb H}}}
\def\CP{{{\Bbb C}{\Bbb P}}}
\def\RP{{{\Bbb R}{\Bbb P}}}
\def\Z{{{\Bbb Z}}}
\def\bA{{{\Bbb A}}}
\def\bB{{{\Bbb B}}}
\def\bC{{{\Bbb C}}}
\def\bR{{{\Bbb R}}}
\def\bD{{{\Bbb D}}}
\def\bE{{{\Bbb E}}}
\def\bZ{{{\Bbb Z}}}
\def\cD{{{\cal D}}}
\def\Re{{{\frak{Re}}}}
\def\Im{{{\frak{Im}}}}
\def\cosec{{\,\hbox{cosec}\,}}
\def\Gm{{\Gamma_{\!\! -}}}
\def\Gp{{\Gamma_{\!\! +}}}

\def\stan{{standard }}
\def\nonstan{{supernumerary }}

\def\cosech{{\hbox{cosech}}}

\def\etcyc{{\hbox{and cyclic}}}
\def\btheta{{\bar\theta}}

\newcommand{\tamphys}{\it Center for Theoretical Physics,
Texas A\&M University, College Station, TX 77843, USA}
\newcommand{\umich}{\it Michigan Center for Theoretical Physics,
University of Michigan\\ Ann Arbor, MI 48109, USA}
\newcommand{\upenn}{\it Department of Physics and Astronomy,\\
University of Pennsylvania, Philadelphia,  PA 19104, USA}
\newcommand{\SISSA}{\it  SISSA-ISAS and INFN, Sezione di Trieste\\
Via Beirut 2-4, I-34013, Trieste, Italy}

\newcommand{\mitchell}{\it George P. \& Cynthia W.
Mitchell Institute for Fundamental Physics,\\
Texas A\&M University, College Station, TX 77843-4242, USA}

\newcommand{\newton}{\it Isaac Newton Institute for Mathematical Sciences,\\
0 Clarkson Road,  University of Cambridge,
Cambridge CB3 0EH, UK}

\newcommand{\ihp}{\it Institut Henri Poincar\'e\\
  11 rue Pierre et Marie Curie, F 75231 Paris Cedex 05}

\newcommand{\damtp}{\it DAMTP, Centre for Mathematical Sciences,
 Cambridge University,\\  Wilberforce Road, Cambridge CB3 OWA, UK}

\newcommand{\itp}{\it Institute for Theoretical Physics, University of
California\\ Santa Barbara, CA 93106, USA}

\newcommand{\istanbul}{\it Department of Mathematics,  Bo{\u g}azi{\c c}i
University, Bebek, Istanbul 34342, Turkey.}

\newcommand{\gursey}{\it Feza Gursey Institute, Cengelkoy, 81220,
Istanbul, Turkey}

\newcommand{\auth}{
G.W. Gibbons\hoch{\star}, R. G\"uven\hoch{\sharp} and
C.N. Pope\hoch{\ddagger}}

\begin{document}
\begin{flushright}
\hfill {
DAMTP-2003-70\ \ \
MIFP-03-17\ \ \
}\\
\hfill{
\bf hep-th/0307238}
\end{flushright}

\begin{center}

{\large {\bf 3-Branes and Uniqueness of the Salam-Sezgin Vacuum}}

\vspace{12pt}

\auth

\vspace{7pt}
 {\hoch{\star}\damtp}

\vspace{7pt}
 {\hoch{\sharp}\istanbul}

\vspace{7pt}
{\hoch{\ddagger}\mitchell}

\vspace{14pt}

\underline{ABSTRACT}
\end{center}

    We prove the uniqueness of the supersymmetric Salam-Sezgin
(Minkowski)$_4\times S^2$ ground state among all nonsingular
solutions with a four-dimensional Poincar\'e, de Sitter or anti-de
Sitter symmetry.  We construct the most general solutions with an
axial symmetry in the two-dimensional internal space, and show
that included amongst these is a family that is non-singular away
from a conical defect at one pole of a distorted 2-sphere.  These
solutions admit the interpretation of 3-branes with negative
tension.

{\vfill\leftline{}\vfill \vskip 10pt \footnoterule {\footnotesize
\hoch{\ddagger} Research supported in part by DOE grant
DE-FG03-95ER40917 \vskip -12pt} \vskip 14pt }
{\footnotesize\hoch{\sharp} Research supported in part by the
Turkish Academy of Sciences (TUBA)}

\pagebreak
\setcounter{page}{1}

\newpage

\section{Introduction}

    There has recently been a revival of interest in the
six-dimensional gauged supergravity model of Salam and Sezgin, which
has long been known to admit a (Minksowki)$_4\times S^2$
supersymmetric vacuum \cite{salsez}, and to have potentially
interesting applications in cosmology
\cite{maenish1,maenish2,gibtown,halliwell,quevedo1,quevedo2}.
On the
theoretical side, it was recently found that this is one of the very
few supergravity models that admits a fully consistent Pauli-type
reduction on a coset space.  Specifically, it was shown that it admits
such a consistent reduction on $S^2$, yielding a chiral four-dimensional $N=1$
supergravity coupled to an $SU(2)$ Yang-Mills multiplet and a scalar
multiplet \cite{gibpop}.  It was also shown that there exists an
extended family of supersymmetric AdS$_3\times S^3$ vacua, with a
parameter characterising the degree of squashing of the $S^3$,
which in an appropriate limit reduce (locally) to the (Minkowski)$_4 \times
S^2$  vacuum \cite{glps}.  On the phenomenological side, the current interest
in large extra dimensions favours six-dimensional models, and the
Salam-Sezgin model has featured in recent studies
(see \cite{quevedo1,quevedo2}, and references therein).

    The Salam-Sezgin model as it stands, being chiral, is anomalous.
These anomalies can be cancelled by the inclusion of additional
matter multiplets \cite{randjbar0,salsez2,bekesase}.  A surprising
feature of the six-dimensional model is that it has a positive
scalar potential and this fact has hindered attempts to obtain it
from higher-dimensional models such as eleven-dimensional
supergravity or ten-dimensional string theory.  Recently, in
\cite{kelu}, it has been shown that the bosonic sector of the
model can be obtained via a generalised dimensional reduction from
$D=7$ and in \cite{cvgp} an M/string-theory origin for the
Salam-Sezgin theory has been found .

   In this letter, we shall show that the remarkable supersymmetric
background found by Salam and Sezgin is in fact unique among all
non-singular backgrounds with four-dimensional Poincar\'e, de Sitter
or anti-de Sitter invariance.  Thus any four-dimensional model based
on the Salam-Sezgin theory must necessarily be supersymmetric unless
3-branes are included, as, for example, introduced in \cite{quevedo2}
by inserting conical defects at the north and south poles of the
2-sphere.  By contrast with many compactifications, such as those
of Calabi-Yau type, which have many moduli coresponding to
flat supersymmetry-preserving directions in the relevant effective
potential, the Salam-Sezgin vacuum has just one free parameter,
which may be taken to be the expectation value of the dilaton field.

   Although the full $SO(3)$ rotational symmetry of the 2-sphere is
broken by the presence of the conical defects in the 3-branes
introduced in \cite{quevedo2}, the solutions are still
axisymmetric. We construct the most general Poincar\'e-invariant
axisymmetric solution, and find that within this class there exist
additional 3-brane solutions (first constructed, in a general
framework, in \cite{gibmae}) with conical defects in which the
local geometry of the 2-sphere is modified from the usual round
$S^2$ geometry, and the dilaton field is no longer constant. The
Einstein equations in these solutions force the existence of
conical defects, without the necessity of introducing additional
delta-function sources in the equations. By contrast with the
3-branes introduced in \cite{quevedo2}, which retain supersymmetry
in the bulk, in our new solutions supersymmetry is broken in the
bulk.

   Unfortunately, the Dirac quantisation
condition forces these branes to have {\it negative} tension.
Following earlier suggestions \cite{randjbar,quevedo2}, one may
incorporate additional six-dimensional gauge fields in the
solutions. These modify the Dirac quantisation condition in a way
which is similar to the modification required for the conical
defects introduced in \cite{quevedo2} but do not alter the sign of
the tension .

   The new 3-brane solutions have a non-constant dilaton field, but are
nevertheless apparently consistent with the suggestion of \cite{quevedo2} that
the 3-brane dilaton coupling should vanish.

\section{Proof of Uniqueness}

    In this section we shall show that any non-singular solution with
a compact internal 2-space and with a four-dimensional spacetime of
maximal symmetry must be the Salam-Sezgin (Minkowski)$_4 \times S^2$
ground state.  We shall do so by first showing that any smooth solution
with compact internal 2-space must be axisymmetric. All axisymmetric
solutions, whether they be singular or not, are then obtained
explicitly.  We then show that the only non-singular solution with
compact internal 2-space in this class is that of  Salam and Sezgin.
 It follows therefore that any smooth ground state  with compact
internal space must be the Salam-Sezgin solution.  Note that we do not
assume axisymmetry; we prove it for all non-singular solutions. Of
course, singular solutions need not be axisymmetric. However, the
explicit axisymmetric (but singular) solutions which we obtain in this
section provide explicit 3-brane solutions whose properties will be
explored in the next section.

    The bosonic sector of the six-dimensional $N=(1,0)$ gauged
Einstein-Maxwell supergravity is described by the Lagrangian
\cite{nissez,salsez}
\be
{\cal L} = \hat R\, {\hat *\oneone} - \ft14 {\hat*d\phi}
\wedge d\phi  - \ft12 e^{\phi}\,
{\hat *H_\3}\wedge H_\3 - \ft12 e^{\ft12\phi}\,
{\hat *F_\2}\wedge F_\2
 - 8g^2 \, e^{-\ft12\phi}\, {\hat *\oneone}\,,\label{d6lag}
\ee
where $F_\2= d A_\1$, $H_\3 = d B_\2 + \ft12
F_\2\wedge  A_\1$, and we place a hat on the six-dimensional metric.
(We use conventions where ${\hat *\omega}\wedge
\omega = (1/p!)\, \omega^{M_1\cdots M_p}\,
\omega_{M_1\cdots M_p}\, {\hat *\oneone}$ for any $p$-form
$\omega$.)  Here $g$ is the gauge-coupling constant, and the
fermions all carry charge $g$ in their minimal coupling to the $U(1)$
gauge field $A_\1$.  The bosonic equations of motion following from
(\ref{d6lag}) are
\bea
\hat R_{MN} &=& \ft14 \del_M\phi\, \del_N\phi +
\ft12 e^{\ft12\phi}\,
(F^2_{MN} - \ft18 F^2\,\hat g_{MN}) + \ft14 e^{\phi}\,
(H^2_{MN} - \ft16 H^2\, \hat g_{MN})\nn\\
&& + 2g^2\,  e^{-\ft12\phi}\, \hat g_{MN}\,,\nn\\
\hat{\nabla}^2\, \phi &=& \ft14 e^{\ft12\phi}\, F^2 +
\ft16 e^{\phi}\, H^2 -
     8g^2\, e^{-\ft12\phi}\,,\nn\\
d(e^{\ft12\phi}\,{*F_\2}) &=& e^{\phi}\,
{*H_\3}\wedge F_\2\,,\label{eoms}\\
d(e^{\phi}\, {*H_\3}) &=& 0\,.\nn
\eea
Note that the dimensionful coupling constant $g$ can be rescaled at
will by adding a constant to $\phi$, together with compensating
rescalings of the other fields \cite{gibpop}.

    It has long been known that this theory admits a solution of the
form (Minkowski)$_4\times S^2$, and furthermore, that this solution
has $N=1$ supersymmetry in the four-dimensional spacetime
\cite{salsez}.  In what follows, we shall demonstrate that the
supersymmetric Salam-Sezgin solution is in fact the only one with
four-dimensional Poincar\'e, de Sitter or anti-de Sitter invariance
and a smooth, non-singular, two-dimensional, compact internal space
$Y$.  We shall do so by first showing that the cosmological constant
for the four-dimensional maximally-symmetric spacetime vanishes.
Then, we shall show that every solution must admit a rotational
Killing vector acting on the internal space, and then we exhibit
explicitly all such axisymmetric solutions.  The only
non-singular one is that of Salam and Sezgin, but there are also
non-supersymmetric solutions with conical singularities, which may be
interpreted as containing 3-branes.  Thus in this case,
non-singularity together with Poincar\'e, de Sitter or anti-de Sitter
invariance implies Poincar\'e supersymmetry, and in order to break
supersymmetry one {\sl must} introduce 3-branes.

   The most general ansatz for a configuration with four-dimensional
 maximal symmetry is
\bea
d\hat s_6^2 &=& W(y)^2 \, ds_4^2 + ds_2^2 \,,\nn\\
H_\3 &=& 0\,,\quad F_{\mu\nu}=0\,,\quad F_{\mu a}=0\,,
\quad F_{ab} = f(y)\, \ep_{ab}\,,
\eea
where $ds_2^2 =g_{mn}\, dy^m\, dy^n$ is the metric on the internal
space $Y$, $W(y)$ is a warp factor, and $ds_4^2$ is a four-dimensional
Minkowski, de Sitter or anti-de Sitter metric.  In the obvious tangent frame,
the components of the six-dimensional Ricci tensor are given by
\be
\hat R_{\mu\nu} = \fft1{W^2}\, R_{\mu\nu}
-\fft1{4W^4}\, \nabla^2 W^4\, \eta_{\mu\nu}\,,\quad
\hat R_{ab} = R_{ab} - \fft4W\, \nabla_a \nabla_b\, W\,,\quad
\hat R_{\mu a}=0\,,
\ee
where $R_{\mu\nu}$ and
$R_{ab}$ are the tangent-frame components of the Ricci tensor
for the four-dimensional spacetime and the internal space, and $\nabla_a$
is the covariant derivative on $Y$.  Our assumptional of maximal
four-dimensional symmetry implies that we shall have
$R_{\mu\nu} = \Lambda\, \eta_{\mu\nu}$.

    The $\hat R_{\mu\nu}$ and $\phi$ equations become, from
(\ref{eoms}),
\bea \ft14  F_\2^2 e^{\ft12\phi}- 8 g^2\, e^{-\ft12\phi} &=&
\fft1{W^4}\,
\nabla^2 W^4 - \fft{4\Lambda}{W^2}\,,\nn\\
\ft14 F_\2^2 e^{\ft12\phi}- 8 g^2\, e^{-\ft12\phi} &=&
\fft1{W^4}\, \nabla^a\, ( W^4\, \nabla_a \phi) \,. \eea
It follows that
\be
\nabla^a\, (W^4\, \nabla_a(\phi - 4\log W))+ 4\Lambda\,
W^2 =0\,.\label{phiw0}
\ee
Integrating over the compact internal manifold $Y$, we immediately see
that $\Lambda\, \int_Y W^2 =0$ and hence the cosmological constant must
vanish.

    Having established that the four-dimensional metric is flat,
we now have
\be
\nabla^a\, (W^4\, \nabla_a(\phi - 4\log W))=0\,.\label{phiw}
\ee
Assuming as before that the internal space $Y$ is complete and
non-singular, and that $\phi$ and $W$ are everywhere smooth
functions on $Y$, with $W$ everywhere positive, we may multiply
equation (\ref{phiw}) by $(\phi - 4\log W)$ and integrate by
parts, to get
\be
\int_Y \sqrt{g}\, d^2y\, W^4\, |\nabla(\phi - 4\log W)|^2 =0\,,
\ee
and hence
\be
\phi = 4 \log W\,.
\ee
(There is no loss of generality in omitting the addititive constant.)
The equation of motion for $F_\2$ now gives
\be
F_\2 = \ft12 q\, W^{-6}\, \ep_{mn}\, dy^m\wedge dy^n \,,
\ee
where $q$ is a magnetic charge.

     Because $Y$ is two-dimensional, we have $R_{mn} = K\, g_{mn}$,
where $K=K(y)$ is the Gauss curvature.  The $R_{mn}$ equation becomes
\be
K\, g_{mn} - \fft2{W^2}\, \nabla_m \nabla_n\, W^2
= \ft38 q^2\, W^{-10}\, g_{mn} + 2 g^2\,
   W^{-2}\, g_{mn}\,.\label{rmneq}
\ee
The tracefree part gives
\be
\nabla_m\nabla_n\, W^2 = \ft12 \nabla^2 W^2\, g_{mn}\,,
\ee
which shows that $\nabla^m\, W^2$ is a conformal Killing vector on
$Y$.  It then follows that
\be
K^m \equiv \ep^{mn}\, \nabla_n\, W^2
\ee
is a Killing vector on $Y$, which is orthogonal to the level sets of
$W$ (and hence $\phi$).

    By integrating the trace of (\ref{rmneq}) over $Y$, one finds that
\be
\chi = \fft1{2\pi}\, \int_Y \sqrt{g}\, K\, d^2y = \fft1{2\pi}\,
\int_Y \sqrt{g}\, d^2y\,
\Big( \fft{4 (\nabla W)^2}{W^2} + \ft38 q^2 W^{-10} + 2 g^2\,
   W^{-2} \Big)\,,
\ee
and hence the Euler number must be positive.  Since we are assuming
that $Y$ is complete, orientable and non-singular, it follows that we
must have $\chi=2$ and $Y$ must be topologically $S^2$.  Moreover, the
Killing vector field $K^m$ must have circular orbits with two fixed
points, that is, $M^m$ is a rotational Killing vector and $Y$ has
axial symmetry.  The most general metric can therefore be written in
the form
\be
d\hat s_6^2 = W^2\, dx^\mu\, dx_\mu + d\rho^2 + a^2\, d\psi^2\,,
\ee
where $W$ and $a$ are functions only of $\rho$.  The equations of motion
then take the form
\bea
\fft{\ddot W}{W} + \fft{3{\dot W}^2}{W^2} + \fft{\dot W\, \dot a}{W\, a}
 &=& \ft14 e^{-\ft12\phi}\, (\ft12 q^2\, W^{-8} - 8 g^2)\,,\nn\\
\fft{4\ddot W}{W} + \fft{\ddot a}{a} + \ft14{\dot\phi}^2 &=&
 -e^{-\ft12\phi}\, (\ft38 q^2 \, W^{-8} + 2g^2)\,,\nn\\
\fft{4\dot W\, \dot a}{W\, a} + \fft{\ddot a}{a} &=&
 -e^{-\ft12\phi}\, (\ft38 q^2 \, W^{-8} + 2g^2)\,,\nn\\
\fft1{a\, W^4}\, \fft{d(a\, W^4\, \dot\phi)}{d\rho}&=&
e^{-\ft12\phi}\, (\ft12 q^2\, W^{-8} - 8 g^2)\,,\label{eomsx}
\eea
where the dot signifies a derivative with respect to $\rho$.  These
equations can be derived from the Lagrangian
\be
L = - 8 W^3\, \dot W\, \dot a - 12a\, W^2\, {\dot W}^2
+\ft14 a\, W^4\, {\dot\phi}^2 - a\, e^{-\ft12\phi}\,
(\ft12 q^2\, W^{-4} + 8 g^2\, W^4)\,,\label{lag1}
\ee
subject to the constraint that the associated Hamiltonian vanishes.

    It follows from (\ref{eomsx}) that there is a constant of the
motion given by
\be
a\, (W^4\, \dot\phi - 4 W^3\, \dot W)= k\,.\label{com}
\ee
As shown above, there are two fixed points of the axial Killing vector
$K^m$ on the smooth $S^2$ manifold, at which the Killing vector field
vanishes.  At these points, 
therefore, $a^2= g_{mn}\, K^m\, K^n=0$.  If we take one of these points,
without loss of generality, to be at $\rho=0$, then 
if $W$ and $\phi$ are smooth functions, bounded at $\rho=0$,
then it is evident that the integration constant $k$ must
vanish.   In appendix A, we construct the most general solutions with
non-vanishing $k$.  Here, we restrict attention to the cases with $k=0$
because, as explained above, only these can give smooth compact
internal spaces.
 
    The local solutions with $k=0$ were written down in
\cite{gibmae}.  They have $\phi=4\log W$, with
\bea
ds_2^2 &=& e^{\ft12\phi}\, \Big( \fft{dr^2}{f_0^2} +
\fft{r^2}{f_1^2}\, d\psi^2 \Big)\,,\nn\\
F_\2 &=& \fft{q\, r}{W^4\, f_0\, f_1}\, dr\wedge d\psi\,,\label{2sol}\\
e^{-\phi} &=& \fft{f_0}{f_1}\,,\qquad
f_0 \equiv 1 + \fft{r^2}{r_0^2}\,,\quad
f_1 \equiv 1 + \fft{r^2}{r_1^2}\,.\nn
\eea
The constants $r_0$ and $r_1$ are given by
\be
r_0^2 = \fft1{2g^2}\,,\qquad r_1^2 = \fft{8}{q^2}\,.\label{r0r1}
\ee

    If $r_1=r_0$, then setting $r=r_0\, \tan \ft12\theta$ one obtains
$W=1$, $\phi=0$ and
\be
ds_2^2 = \ft14 r_0^2\, (d\theta^2 + \sin^2\theta\, d\psi^2)\,,
\ee
which is the round $S^2$ metric of the Salam-Sezgin solution.  As we shall
see in detail in the next section, this is the only completely regular
solution.  Our proof of the uniqueness is thus complete.

\section{3-Brane Solutions}\label{3bsec}

    When $r_0 \ne r_1$, one finds that if $\psi \in [0,2\pi)$, then $Y$
is smooth at $r=0$ but has a conical singularity at $r=\infty$, with
deficit angle $\delta$ given by
\be \fft{\delta}{2\pi} = 1 - \fft{r_1^2}{r_0^2}\,. \ee
This conical singularity represents a 3-brane with positive
tension if $r_0 > r_1$, and negative tension if $r_0 < r_1$.  The
field $F_\2$ can be written locally in terms of the 1-form
potential
\be
A_\1 = - \fft{4}{q\, f_1}\, d\psi\,.
\ee
This is well-behaved as $r$ goes to infinity, but not at the origin.
Performing the gauge transformation $A_1 \longrightarrow A_1
+ d(4\psi/q)$ gives a potential which is regular near the origin, and so
single-valuedness of the fermionic fields requires that
the Dirac quantisation condition
\be
\fft{4 g}{q} = N
\ee
must be satisfied, where $N$ is an integer.  Equivalently, the flux
\be
\fft{1}{4\pi} \, \int_Y F_\2 =\fft{2}{q}
\ee
is quantised in units of $1/(2g)$.

   From (\ref{r0r1}) it follows that the deficit angle at $r=\infty$
is given by
\be \fft{\delta}{2\pi} = 1 - N^2, \ee
and that the ratio $r_1/r_0$ is quantised
\be
\fft{r_1}{r_0} = |N|\,.
\ee
Unfortunately, this implies for $\mid N \mid > 1$ that the 3-brane
tension is necessarily negative.

More generally, one may identify $\psi $ with period $2 \pi \alpha$, 
where $\alpha >0$. The deficit angle is given by 
\be
\delta = 2 \pi - \lim_{ \rho \rightarrow 0}   {C(\rho ) \over \rho},  
\ee
where $C(\rho)$ is the circumference of a small circle of radius $\rho$.
Thus at $r=0$ and $r= \infty$ the deficits are 
\be
\delta_0= 2 \pi (1-\alpha) \,, \qquad \delta_\infty = 2 \pi 
( 1- {N^2 \over \alpha })\,.
\ee
The tension is given in terms of the deficit angle by
\be
T= { \delta \over 8  \pi G},
\ee
which implies
\be
T_0= { 1 \over 4G} \, (1-\alpha) \,, \qquad T_\infty = { 1 \over 4G}
\, ( 1- {N^2 \over \alpha })\,.
\ee
Thus both $T_0$ and $T_\infty$ are less than $1 \over 4G$, and 
\be
(1-4G\, T_0)(1-4G\, T_\infty) = N^2\,.
\ee
If the integer $N$ exceeds $1$, then it follows that both tensions, 
$T_0$ and $T_\infty$, must be negative.

\section{Solutions With Additional Gauge Fields}

    In \cite{quevedo2}, following earlier work of \cite{randjbar}, the
2-form supporting the solution was taken to be a linear
combination of the supergravity 2-form $F_\2$ that we have been using thus
far, and a $U(1)$ subgroup of an additional Yang-Mills gauge sector $F^I_\2$
in the six-dimensional theory.  Thus now
\be
F_\2 = \fft{q\, r\, \cos\beta}{W^4\, f_0\, f_1}\, dr\wedge d\psi
\,,\qquad
T_I\, F^I_\2 = T_0\,  \fft{q\, r\, \sin\beta}{W^4\, f_0\, f_1}\, dr\wedge d\psi
\,,
\ee
where $\beta$ is the mixing angle, and $T_0$ denotes the $U(1)$ generator
within the Yang-Mills sector.  There are now two Dirac quantisation
conditions, associated with the requirement of single-valuedness for the
supergravity and gauge-sector fermions respectively:
\be
\fft{4 g\, \cos\beta}{q} = N\,,\qquad
\fft{4g'\, \sin\beta}{q} = N'\,,
\ee
where $g'$ is the relevant gauge coupling constant in the Yang-Mills sector,
and $N$ and $N'$ are integers.

     Using (\ref{r0r1}), we can re-express these conditions as
\be
\fft{r_1}{r_0} = \fft{N}{\cos\beta}\,,\qquad
\fft{g'}{g} = \fft{N'}{N}\, \cot\beta\,.
\ee
The first equation can always be solved, provided that $r_1>r_0$,
which implies as  before that the 3-brane will not have a positive
tension.  The second equation may then be regarded as determining
$g'$.  Note that these Dirac quantisation conditions are similar
to those obtained in \cite{gibpop}, where, following
\cite{quevedo2}, conical deficits $2\pi\, \epsilon$ were
introduced at the north and south poles of a round $S^2$.  In that
case, the analogous quantisation conditions were \cite{gibpop}
\be
\cos\beta = \fft{N}{1-\epsilon}\,,\qquad
\fft{g'\, \sin\beta}{g} = \fft{N'}{1-\epsilon}\,.\label{tw2}
\ee
The special cases $\beta=0$ and $\beta=\ft12\pi$ were obtained
earlier in \cite{quevedo2}. It was noted in \cite{gibpop} that the
first equation in (\ref{tw2}) could not be satisfied for any
integer $N$ when $|\cos\beta|\ne 1$ or 0, unless $\epsilon$ was
taken to be negative; in other words the 3-brane tension had to be
negative.

\section{3-Brane/Dilaton Coupling}

    In \cite{quevedo2}, 3-branes were introduced into the Salam-Sezgin
model by inserting conical
deficits at the north and south poles of the 2-sphere, with the
dilaton being independent of the coordinates on $S^2$.  The 3-brane
action was taken to be
\be
S_b = -T\, \int d^4x\, e^{-\ft12 \lambda\, \phi}\,
(-\det \gamma_{\mu\nu})^{1/2}\,, \label{3brane}
\ee
where $\gamma_{\mu\nu}= \hat g_{MN}\, \del_\mu X^M\, \del_\nu X^N$ is the
induced metric on the 3-brane.\footnote{Our $\phi$ is $(-2)$ times
the $\phi$ in \cite{quevedo2}, and so $\lambda$ is the same as that used in
\cite{quevedo2}.}  In the detailed calculations in \cite{quevedo2}, the
3-brane/dilaton coupling $\lambda$ was taken to be zero.

   In the more general solutions (\ref{2sol}) obtained in this paper,
3-branes arise naturally when $r_1\ne r_0$.  In these solutions
the dilaton is not constant, and this allows us to make
qualitative statements about the 3-brane/dilaton coupling.  For
negative-tension 3-branes, \ie $r_1 > r_0$, the dilaton decreases
from its value at the origin as one aproaches the 3-brane at
$r=\infty$.  Conversely, if the tension is positive, \ie $r_1 <
r_0$, the dilaton increases as the 3-brane at $r=\infty$ is
approached. The fact that in our solutions $\phi$ is a smooth
function without singularities is consistent with the idea that
the 3-brane/dilaton coupling $\lambda$ is in fact zero, as
proposed in \cite{quevedo2}, because otherwise one would expect
singular behaviour near the 3-brane from the delta-function in the
dilaton equation arising from the contribution (\ref{3brane}) to
the action.

\section{Modulus and Breathing Mode}

   Our proof of uniqueness shows that the Salam-Sezgin ground state
has just one modulus, namely the value of $\phi_0$.  One can consider
solutions in which the radius of the 2-sphere varies in space and time,
with the six-dimensional fields taking the forms
\bea
d\hat s_6^2 &=& e^{\ft12(\phi_1+\phi_2)}\, ds_4^2 +
e^{-\ft12(\phi_1+\phi_2)}\, g_{mn}\, dy^m\, dy^n\,,\nn\\
F_\2 &=& 4g\, \ep_\2\,,\qquad \phi= \phi_2 -\phi_1\,,\qquad
H_\3=0\,,
\eea
where $\ep_\2$ is the volume-form of the metric $g_{mn}\, dy^m\, dy^n$
on $S^2$, which we normalise to $R_{mn}= 8g^2\, g_{mn}$.
Substituting into the higher dimensional action, which is a valid
procedure since this dimensional reduction is trivially consistent,
yields the four-dimensional action
\be
{\cal L} = R  - \ft12 (\del\phi_1)^2 - \ft12
(\del\phi_2)^2 - 8g^2\, e^{\phi_1}\, (1-e^{\phi_2})^2\,.
\label{2scal}
\ee
The potential in (\ref{2scal}) was first derived, in the purely time-dependent
case, in \cite{maenish1}, and some cosmological applications were given in
\cite{maenish1,maenish2,gibtown,halliwell}.

    The field $\phi_2$ plays the role of a breathing mode (or ``radion'').
Its mass $M_{\rm KK}$ is given by
\be
M_{\rm KK} = 4g\, e^{\ft12\phi_0}\,,
\ee
where $\phi_0$ denotes the expectation value of the massless
``modulus scalar'' $\phi_1$.  As pointed out in \cite{gibpop}, all
Kaluza-Klein modes have masses set by the mass of this radion
field.

\pagebreak

\section*{Acknowledgments}

    We are grateful to Cliff Burgess, Jim Liu, Hong L\"u, Fernando
Quevedo and Ergin Sezgin for discussions.
We are especially grateful to Seif Randjbar-Daemi for pointing out 
an error in the sign of the brane tension in an earlier version of
this paper.  G.W.G. and C.N.P. are grateful to the Feza Gursey Institute,
Istanbul, and the Benasque Center for Science, for hospitality
during the course of this work.

\appendix
\section{General Axisymmetric Solutions}

    Here we construct the most general solution to the equations (\ref{eomsx})
for axially-symmetric configurations.  It is advantageous first to introduce
the ``lapse function'' ${\cal N}$ in the Lagragian (\ref{lag1}), which enforces
the vanishing of the associated Hamiltonian:
\be
L = (- 8 W^3\, \dot W\, \dot a - 12a\, W^2\, {\dot W}^2
+\ft14 a\, W^4\, {\dot\phi}^2)\, {\cal N}  - a\, e^{-\ft12\phi}\,{\cal N}^{-1}
(\ft12 q^2\, W^{-4} + 8 g^2\, W^4)\,,\label{lag2}
\ee
We next send ${\cal N} \longrightarrow {\cal N}/(a\, W^4)$,
make the coordinate gauge transformation $d\rho = a\, W^4\, d\eta$, and
then suppress the lapse function.  After introducing new independent
variables by defining
\be
W = e^{\ft14(y-x)}\,,\qquad a = e^{\ft14(3x+y+2z)}\,,\qquad
\phi = y-x + 2z\,,
\ee
we obtain the Lagrangian
\be
{x'}^2 - {y'}^2 + {z'}^2 - \ft12 q^2\, e^{2x} + 8g^2\, e^{2y}\,,
\label{lag3}
\ee
together with the Hamiltonian constraint
\be
{x'}^2 - {y'}^2 + {z'}^2 + \ft12 q^2\, e^{2x} - 8g^2\, e^{2y}=0\,,
\label{hamil}
\ee
where a prime denotes a derivative with respect to $\eta$.

    In terms of the new variables, the general system of equations
of motion is decoupled, reducing to two Liouville equations for $x$
and $y$, and a free-particle equation for $z$.  We have the three first
integrals
\be
{x'}^2 + \ft12 q^2\, e^{2x}\, =\lambda_1^2\,,\qquad
{y'}^2 + 8g^2\, e^{2y} = \lambda_2^2\,,\qquad
z' = \lambda_3\,,
\ee
and the Hamiltonian constraint implies that the three constants of integration
obey the relation
\be
\lambda_2^2 = \lambda_1^2 + \lambda_3^2\,.
\ee
Note that $\lambda_3$ is related to the constant $k$ in (\ref{com}) by
$k=2\lambda_3$.

The general solution can be taken, without loss of generality, to be given by
\be
e^{-x} = \fft{q}{\sqrt2\, \lambda_1}\, \cosh\lambda_1\, (\eta-\eta_1)
\,,\qquad
e^{-y} = \fft{2 \sqrt2\, g}{\lambda_2}\,
\cosh\lambda_2\, (\eta-\eta_2)
\,,\qquad
z = \lambda_3\, \eta\,,
\ee
The metric and dilaton are therefore given by
\bea
d\hat s_6^2 &=& W^2\, dx^\mu\, dx_\mu + a^2\, W^8\, d\eta^2 + a^2\, d\psi^2
\,,\nn\\
e^\phi &=& W^4\, e^{2\lambda_3\, \eta}\,,
\eea
where $W$ and $a$ are given by
\bea
W^4 &=& \fft{q\, \lambda_2}{4g\, \lambda_1}\,
\fft{\cosh\lambda_1(\eta-\eta_1)}{\cosh\lambda_2(\eta-\eta_2)}\,,\nn\\
a^{-4} &=& \fft{g\, q^3}{\lambda_1^3\, \lambda_2} \, e^{-2\lambda_3\, \eta}\,
\cosh^3\lambda_1(\eta-\eta_1)\, \cosh\lambda_2(\eta-\eta_2)\,.
\eea

   The solutions in section 2 that are regular at the origin correspond to
taking $\lambda_3=0$, and hence $\lambda_1=\lambda_2$.  This solution, in the
form (\ref{2sol}), is obtained by setting
\be
\lambda_1=\lambda_2=1\,,\qquad r = r_1\, e^{\eta-\eta_1}\,,\qquad
e^{\eta_1-\eta_2} = \fft{4g}{q}\,.
\ee



\begin{thebibliography}{99}

\bm{salsez}
A. Salam and E. Sezgin,
{\it Chiral compactification on (Minkowski)$\times S^2$ of $N=2$
Einstein-Maxwell supergravity in six-dimensions},
Phys. Lett. {\bf B147}, 47 (1984).

\bibitem{maenish1}
K.I.~Maeda and H.~Nishino,
{\it Cosmological solutions in $D=6$, $N=2$ Kaluza-Klein supergravity:
Friedmann universe without fine tuning},
Phys.\ Lett. {\bf B154}, 358 (1985).

\bibitem{maenish2}
K.I.~Maeda and H.~Nishino,
{\it Attractor universe in six-dimensional $N=2$ supergravity Kaluza-Klein
theory},
Phys.\ Lett. {\bf B158}, 381 (1985).

\bm{gibtown} G.W. Gibbons and P.K. Townsend,
{\it Cosmological evolution of degenerate vacua},
Nucl. Phys. {\bf B282}, 610 (1987).

\bibitem{halliwell}
J.J.~Halliwell,
{\it Classical and quantum cosmology of the Salam-Sezgin model},
Nucl.\ Phys. {\bf B286}, 729 (1987).

\bibitem{quevedo1}
Y. Aghababaie, C.P. Burgess, S.L. Parameswaran and F. Quevedo,
{\it SUSY breaking and moduli stabilization from fluxes in gauged 6D
supergravity},
JHEP {\bf 0303}, 032 (2003), hep-th/0212091.

\bibitem{quevedo2}
Y.~Aghababaie, C.P.~Burgess, S.L.~Parameswaran and F.~Quevedo,
{\it Towards a naturally small cosmological constant from branes in 6D
supergravity},
hep-th/0304256.

\bm{gibpop} G.W. Gibbons and C.N. Pope,
{\it Consistent S**2 Pauli reduction of six-dimensional chiral gauged
Einstein-Maxwell supergravity}, hep-th/0307052.

\bm{glps} R. G\"uven, J.T. Liu, C.N. Pope and E. Sezgin, {\it Fine
tuning and six-dimensional gauged $N = (1,0)$ supergravity vacua},
Class. Quantum Grav. {\bf 21}, 1001 (2004), hep-th/0306201.

\bm{randjbar0} S. Randjbar-Daemi, A. Salam, E. Sezgin and J.
Strathdee, {\it An anomaly free model in six-dimensions}, Phys.
Lett. {\bf B151}, 351 (1985).

\bm{salsez2}
A.~Salam and E.~Sezgin,
{\it Anomaly Freedom In Chiral Supergravities},
Phys.\ Scripta {\bf 32}, 283 (1985).

\bibitem{bekesase}
E.~Bergshoeff, T.~W.~Kephart, A.~Salam and E.~Sezgin,
{\it Global Anomalies In Six-Dimensions},
Mod.\ Phys.\ Lett. {\bf A1}, 267 (1986).

\bm{kelu} J. Kerimo and H. L\"u, {\it New $D=6$, $N=(1,1)$ gauged
supergravity with supersymmetric (Minkowski)$_4 \times S^2$
vacuum}, hep-th/0307222.

 \bm{cvgp} M. Cveti$\breve{c}$, G.W. Gibbons and
C.N. Pope, {\it A string and M-theory origin for the Salam-Sezgin
model} Nucl. Phys. {\bf B677} 164, (2004), hep-th/0308026.

 \bm{gibmae} G.W. Gibbons and K.I.
Maeda, {\it Black holes and membranes in higher dimensional
theories with dilaton fields}, Nucl.\ Phys. {\bf B298}, 741
(1988).


\bm{randjbar} S. Randjbar-Daemi, A. Salam and J. Strathdee,
{\it Spontaneous compactification in six-dimensional Einstein-Maxwell theory},
Nucl. Phys. {\bf B214}, 491 (1983).

\bibitem{nissez} H.~Nishino and E.~Sezgin,
{\it Matter and gauge couplings of $N=2$ supergravity in six-dimensions},
Phys.\ Lett. {\bf B144}, 187 (1984).


\end{thebibliography}
\end{document}